\pdfoutput=1

\documentclass[fleqn,10pt]{wlscirep}
\usepackage[utf8]{inputenc}
\usepackage[T1]{fontenc}
\usepackage{color,soul}
\usepackage{float}
\usepackage{enumitem}
\usepackage{amsmath}
\usepackage{array}
\usepackage{epstopdf}
\title{Travel experience in public transport: Experience sampling and cardiac activity data for spatial analysis}

\author[1,*]{Esther Bosch}
\author[2]{Ricarda Luther}
\author[1]{Klas Ihme}
\affil[1]{German Aerospace Center, Institute of Transportation Systems, Braunschweig, 38106, Germany}
\affil[2]{Institute for Information Management, Bremen, 28359, Germany}

\affil[*]{esther.bosch@dlr.de}

\begin{abstract}
The transportation sector has the potential to enable a greener future if aligned with increasing mobility needs. Making public transport an attractive alternative to individual transportation requires real-world data to investigate reasons and indicators of positive and negative travel experiences. These experiences manifest not only in subjective evaluations but also in physiological reactions like cardiac activity. We present a geo-referenced dataset where participants wore electrocardiograms and reported real-time stress, satisfaction, events, and emotions while traveling by tram, train, and bus. An interactive experience map helps to visually explore the data, with benchmark analyses identifying significant stress hot spots and satisfaction cold spots during journeys. Events and emotions in these spots highlight positive and negative travel experiences in an ecologically valid setting. Data on age and self-identified gender provide insights into differences between user groups. Despite including only 44 participants, the dataset offers a valuable foundation for transportation researchers and mobility providers to combine qualitative and quantitative methods for identifying public transportation users' needs.
\end{abstract}
\begin{document}

\flushbottom
\maketitle

\thispagestyle{empty}

%\noindent Please note: Abbreviations should be introduced at the first mention in the main text – no abbreviations lists or tables should be included. Structure of the main text is provided below.

\section*{Background \& Summary}

Transport accounts for nearly 25\% of global carbon dioxide emissions from energy use \cite{owid-travel-carbon-footprint}. In the EU, buses, railways, and trams made up 13.8\% of motorized kilometers traveled on land in 2020, while cars dominated with 86.2\% of kilometers driven \cite{eubook}. Public transport, however, produces more than 50\% less greenhouse gas emissions per passenger kilometer compared to small and medium-sized cars \cite{IEA}. Shifting from private to public transport is thus a critical step toward reducing global greenhouse gas emissions \cite{ipcc2022}.

While the need to increase public transport usage is well-researched, the feasibility of this shift has been extensively studied, with research focusing on factors such as efficiency \cite{cortes2011integrating, basso2011congestion}, infrastructure \cite{yang2016incorporating}, and policy interventions \cite{galilea2010does}. Another important factor in this transition is the user, whose preferences and decisions ultimately drive public transport adoption. Previous studies have primarily addressed practical considerations like time, cost, and reliability that influence mobility choices \cite{Schiefelbusch.2015, Chowdhurry2015, schakenbos2016}.

However, beyond these practical considerations, the emotional and experiential dimensions of travel remain underexplored. These dimensions are critical, as they influence user satisfaction and long-term loyalty to public transport. Lim et al. \cite{lim2024effects} introduce the concept of within-trip subjective experiences — cognitive and affective responses to individual travel stimuli mediated by appraisal processes — and propose that these experiences shape travel satisfaction, influence mode-related attitudes and intentions, and potentially modify underlying beliefs about mode choice. Similarly, De Vos et al. \cite{de2022attitude} demonstrate that attitudes are strongly linked to future travel mode choices. However, existing research on travel experience has largely relied on retrospective questionnaires when inquiring about within-trip subjective experiences. The Satisfaction with Travel Scale (STS) \cite{ettema2011satisfaction} is one such instrument, designed to measure two affective and one cognitive component of travel experience \cite{chatterjee2020commuting}. Olsson et al. \cite{olsson2012measuring} were among the first to apply the STS to public transport, analyzing 361 valid questionnaires. Their findings confirmed the factor structure of the STS, with two affective and one cognitive dimension, achieving a Cronbach’s alpha of 0.78. Lunke \cite{Lunke2020} extended this work by analyzing responses from 14,015 participants using a three-item questionnaire derived from the STS. This study revealed that short waiting times and reliable transport are key drivers of travel satisfaction.

In addition to the STS, alternative methods have been employed to investigate travel experience. Beck et al. \cite{beck2016best} developed a best-worst scaling questionnaire to assess attitudes toward aspects of bus trips, including noise levels and the availability of information. Soza Parra et al. \cite{soza2019underlying} implemented an in-situ measurement approach, assessing satisfaction per travel leg on a seven-point scale at four metro stations in Santiago de Chile. This real-time approach reduces reliance on retrospective recollection. Meanwhile, Van Lierop \cite{Lierop.2017} conducted a literature review to identify factors influencing public transport loyalty, highlighting attributes such as cleanliness, comfort, operator helpfulness, and schedule frequency. Ingvardson et al. \cite{ingvardson2019relationship} analyzed 44,956 questionnaires from six European cities using structural equation modeling, identifying accessibility, comfort, staff behavior, and safety/security as the top four dimensions affecting public transport satisfaction.

Despite these advancements, the research remains limited in scope, particularly regarding real-time, multimodal investigations of travel experience. Travel experiences are inherently tied to emotions, which involve both cognitive appraisals and physiological responses \cite{scherer2009dynamic}. Retrospective methods often fail to capture the dynamic and situational aspects of these experiences, as they are subject to memory biases that can distort both the intensity and frequency of reported events \cite{redelmeier1996patients}. Experience sampling offers a promising alternative, enabling the collection of subjective data with high temporal resolution in ecologically valid settings \cite{csikszentmihalyi2014validity}. This method minimizes memory bias by capturing moments of positive and negative experiences as they occur, providing richer and more accurate insights compared to post-hoc approaches \cite{scollon2003experience}.

Physiological data, such as cardiac activity as measured by heart rate and variability, can complement subjective reports by providing objective indicators of emotional states. Previous studies have successfully correlated cardiac activity changes with subjective experiences and emotions \cite{saganowski2022emognition, shui2021dataset, shu2018review, brouwer2018improving, brandebusemeyer2022travelers}. Especially Heart Rate Variability (HRV) has been found to correlate with sympathetic nervous system activity \cite{rajendra2006heart, electrophysiology1996heart} and therefore higher arousal emotions \cite{shi2017differences}, cortical arousal \cite{basner2007ecg}, and stress \cite{kim2018stress}. Castro et al. \cite{castro2020methodological} incorporated cardiac activity measurements into a model of travel experience, presenting a case study with one participant to demonstrate the feasibility of such methods. Additionally, location-based visualizations of experience sampling data, such as maps, have been used to analyze collective travel experiences. These methods have been applied in contexts like automotive travel \cite{Dittrich.2021, Greco.2015} and pedestrian or cyclist experiences \cite{kyriakou2019detecting}. Bosch et al. \cite{Bosch_maps} proposed using experience mapping to investigate public transport experiences, suggesting its potential to pinpoint collective pain points in transportation systems.

Despite the recognized importance of emotions and experiences in shaping public transport satisfaction, existing research relies heavily on retrospective and subjective data collection methods. These methods are prone to memory biases and fail to capture the dynamic, real-time aspects of travel experiences. Furthermore, the integration of subjective, cardiac activity, and contextual data in public transport research remains underexplored. This paper addresses these gaps by presenting a comprehensive dataset collected from 44 participants who traveled on a predefined 15 km intermodal route involving tram, bus, and train transitions. The dataset includes global positioning system (GPS) data, heart rate and heart rate variability measurements, and real-time experience sampling data on travel satisfaction, stress levels, self-reported emotions, and significant events. The data is visualized in an interactive map, enabling the identification of collective hot spots of stress and satisfaction and the associated events and emotions. By integrating subjective, cardiac activity, and contextual dimensions, this study provides a synchronous and holistic approach to understanding travel experiences.

This dataset facilitates the development of methods for analyzing the causes of positive and negative travel experiences, identifying collective pain points, and investigating cardiac activity correlates of experiences during intermodal transportation. Additionally, demographic data on age and self-identified gender supports analyses of group-specific differences, enabling a refined understanding of user needs in public transportation.

\section*{Methods}

%The Methods should include detailed text describing any steps or procedures used in producing the data, including full descriptions of the experimental design, data acquisition assays, and any computational processing (e.g. normalization, image feature extraction). See the detailed section in our submission guidelines for advice on writing a transparent and reproducible methods section. Related methods should be grouped under corresponding subheadings where possible, and methods should be described in enough detail to allow other researchers to interpret and repeat, if required, the full study. Specific data outputs should be explicitly referenced via data citation (see Data Records and Citing Data, below).

%Authors should cite previous descriptions of the methods under use, but ideally the method descriptions should be complete enough for others to understand and reproduce the methods and processing steps without referring to associated publications. There is no limit to the length of the Methods section. Subheadings should not be numbered.

\subsection*{Procedure}
\begin{figure}[ht]
\centering
\includegraphics[scale=0.8]{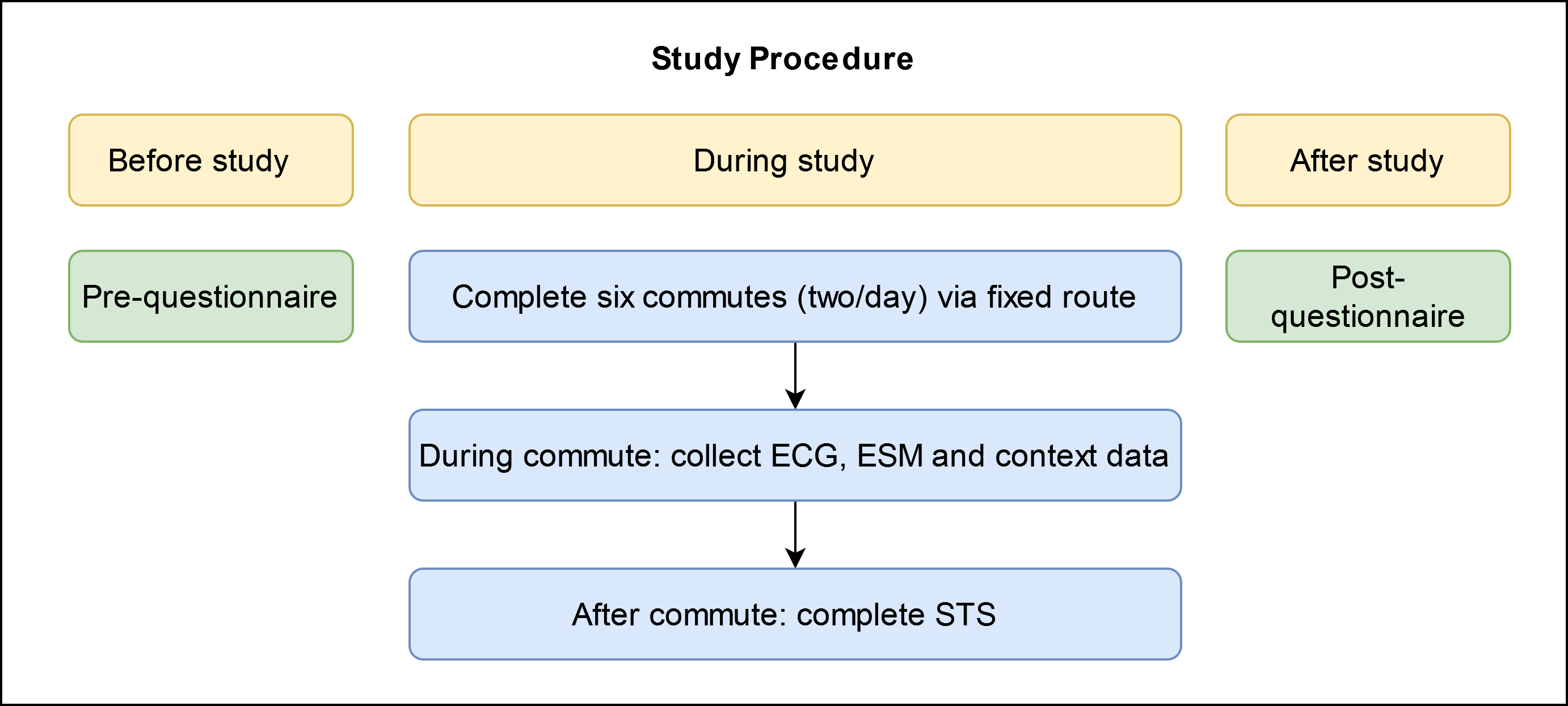}
\caption{Study procedure.}
\label{fig:procedure}
\end{figure}

Participants arrived at our institute and were informed about our study's procedure (see \autoref{fig:procedure}) and aim. They were equipped with a cell phone and an electrocardiogram (ECG) belt, which they could keep for up to three weeks. Within this time, all participants traveled the same predefined route in both directions between three to five times, depending on the participant's availability. They recorded their way from and to the research institute, where they were free to choose a transportation mode. The one-way route consisted of 

\begin{figure}[ht]
\centering
\includegraphics[width=0.35\linewidth]{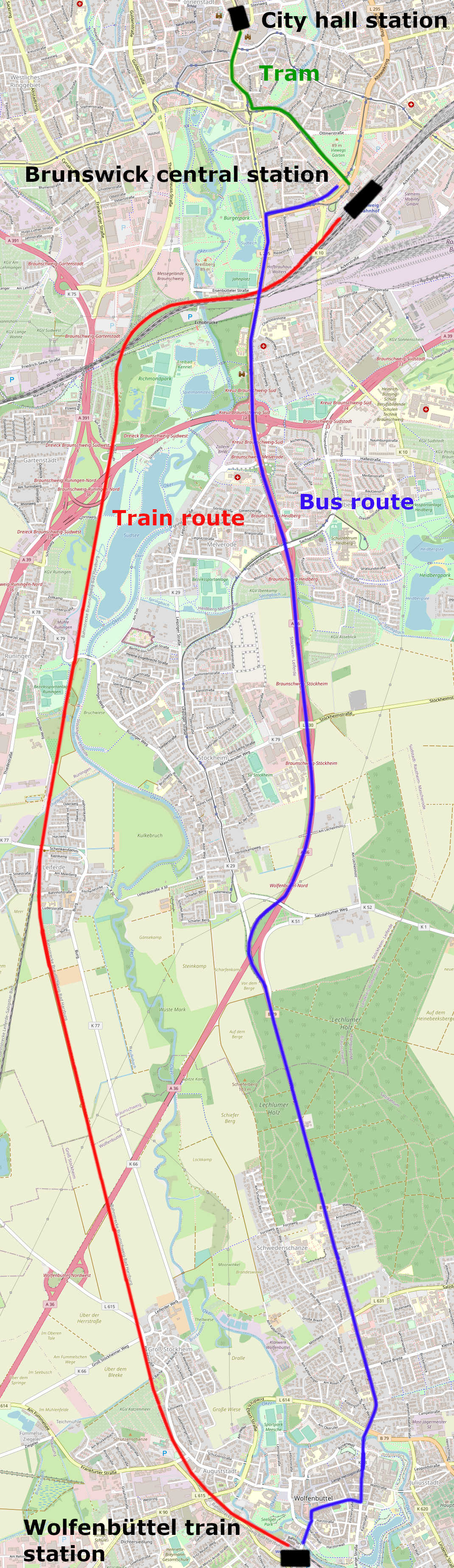}%screenshot rshiny
\caption{Travel modes on the route.}
\label{travelmodes}
\end{figure}

\begin{itemize}[itemsep=0.5pt]
    \item Getting from home with a freely chosen transportation mode to a central bus station in Braunschweig (German city with roughly 250,000 inhabitants),
    \item taking a tram to Braunschweig main station (GPS location: 52.251332328, 10.537164518), 
    \item taking a bus (50\% of journeys) or a train (the other 50\% of journeys) to Wolfenbüttel (German city with roughly 50,000 inhabitants, distance roughly 15km, GPS location: 52.15951774394025, 10.532477626246047), 
    \item walking to the city center (Schlossplatz Wolfenbüttel, GPS location: 52.163224559025245, 10.531477397957316).
\end{itemize}

and was taken back in reverse. Participants were free to choose to start their journey from Braunschweig or Wolfenbüttel. The route was chosen to include as many as possible transportation modes of public transport. \autoref{travelmodes} displays the route. Data acquisition took place from January to December 2023. 

Figures $bs_wf_route.png, 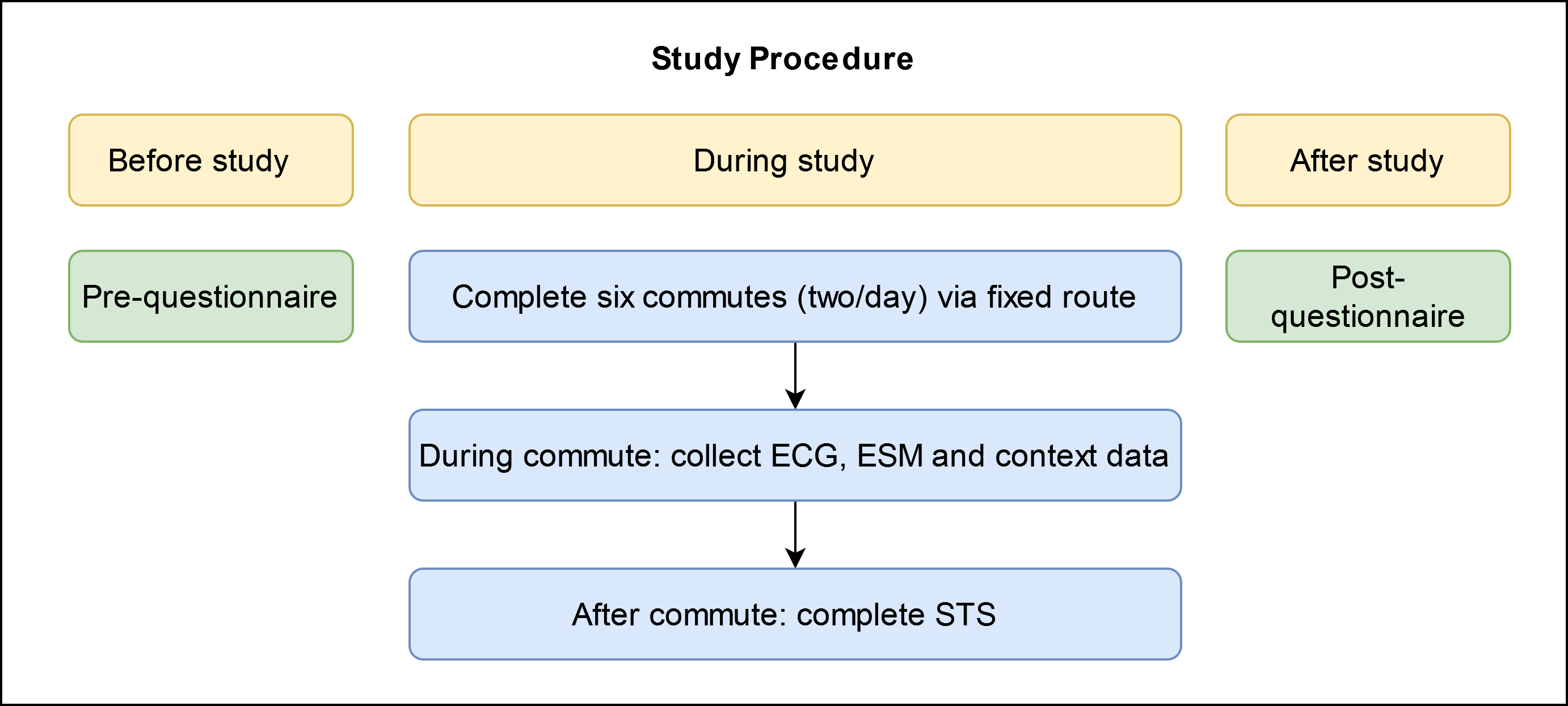$ go here

\subsection*{Participants}
In total, 44 participants over the age of 18 years (23 self-identified as male, 21 self-identified as female) took part in the study. The participants were recruited from our institute's participant pool, which includes individuals who have previously signed up to participate in studies. To mitigate potential recruitment biases, we ensured that the sample was balanced by gender. Participants were selected based on their availability and willingness to participate, and no specific demographic quotas beyond gender balance were applied. Their commute distances, size of the city they live in, education levels and commute transport modes, educational background are shown in Tables \ref{tab:commutedistance}, \ref{tab:clivingsituation}, \ref{tab:education}, and \ref{tab:commutemode}. Their answers to the Big Five Inventory (BFI-10), and to the Affinity to Technology (ATI-9) scale are provided in the columns 'BFI' and 'ATI'. They had a mean age of 29 years (SD 14 years), and none of them indicated to have had any cardiovascular diseases. They indicated to have no specific restrictions during travel except for two persons who indicated to travel with a lot of luggage on all of their journeys. The participants received 60€-100€ (depending on the number of journeys taken, 10€ per journey) as financial reimbursement for participation. Participants who did not hold a yearly regional public transport ticket were compensated for their travel expenses during the study. Before starting the study, participants were informed about and agreed to the study's procedure and the handling of personal data (in line with the General Data Protection Regulation of the EU, approved by the institution's data protection and legal committee). The study's procedure was approved by the institution's ethical committee (reference no. 07/22).

Tables 1-4 go here

\subsection*{Measurements}\label{methods}

In this study, subjective travel experience was assessed across three distinct time frames: (1) once per participant (e.g., demographic data, personality traits measured by the Big Five Inventory, and technology affinity via the Affinity for Technology Interaction scale); (2) once per journey (e.g., destination details and an overall satisfaction evaluation); and (3) during travel, using experience sampling questionnaires administered at 3.5-minute intervals. This multi-method approach integrates real-time data with post-travel assessments and mode choice questionnaires, enabling an investigation of decision-making heuristics \cite{redelmeier1996patients} and their relationship with during-travel experiences, post-journey satisfaction with travel (STS), and future mode choice behaviors.

This methodology combines the strengths of experience sampling and retrospective questionnaires. Experience sampling enhances ecological validity, captures within-person variability, and reduces memory bias \cite{scollon2003experience}. On the downside, the highly regular time intervals of reporting experience will make participants more aware of their internal states, which in turn might change them  \cite{wheeler1991self}. In contrast, retrospective questionnaires avoid repeated assessments and therefore minimize atypical focus on internal states and provide a broader contextual overview \cite{scollon2003experience}. Furthermore, the inclusion of physiological data adds an objective dimension to the subjective real-time experience, capturing multimodal aspects of travel experiences, as proposed by Scherer’s dynamic framework \cite{scherer2009dynamic}. These physiological measures complement self-report data, which can be constrained by social desirability, cognitive biases, cultural norms, and the limitations of numerical scales \cite{scollon2003experience}. 

Contextual factors like time of day and year, weather and delays significantly influence travel satisfaction, shaping both real-time experiences and post-travel attitudes. With the provided GPS and timestamp data, it is possible to obtain weather information and delays. Drawing on Kaiser’s framework \cite{kaiser1999environmental}, which emphasizes the interaction of attitudes, behavior intentions, and external constraints, these situational factors mediate the relationship between during-travel experiences, satisfaction, and future transport choices. Physiological stress markers (e.g., heart rate) reveal how these stressors manifest physically, while subjective self-reports capture individual perceptions influenced by personal attitudes. Integrating real-time contextual data, such as the weather, with subjective and physiological measures provides a comprehensive understanding of how external constraints impact travel experiences.

\subsubsection*{Subjective measures - once per participant:} Before the start of the study, participants filled in a demographic questionnaire indicating their usual commute distance and mode, the size of the city they live in and education level. They also answered the Big Five Inventory (BFI-10) following \cite{rammstedt2013short} on a scale from 1 = strongly disagree to 5 = strongly agree with the following questions:
\begin{itemize}[itemsep=0.5pt]
    \item "I am rather reserved" - scores on Extraversion, reversed
    \item "I am generally trusting" - scores on Agreeableless
    \item  "I tend to be lazy" - scores on Conscientiousness, reversed
    \item  "I am relaxed and handle stress well" - scores on Neuroticism, reversed
    \item  "I have few artistic interests" - scores on Openness, reversed
    \item "I am outgoing, sociable" - scores on Extraversion
    \item  "I tend to find fault with others" - scores on Agreeableless, reversed
    \item  "I do a thorough job" - scores on Conscientiousness
    \item  "I get nervous easily" - scores on Neuroticism
    \item  "I have an active imagination" - scores on Openness

\end{itemize}
They also filled in the Affinity to Technology (ATI-9) scale following \cite{franke2019personal} that is computed as mean over all 9 scores on a scale from 1 = 'completely disagree' to 6 = 'completely agree' with the following questions:
\begin{itemize}[itemsep=0.5pt]
    \item "I like to occupy myself in greater detail with technical systems"
    \item "I like testing the functions of new technical systems"
    \item  "I predominantly deal with technical systems because I have to" - reversed
    \item  "When I have a new technical system in front of me, I try it out
intensively"
    \item  "I enjoy spending time becoming acquainted with a new
technical system"
    \item "It is enough for me that a technical system works; I don’t care
how or why" - reversed
    \item  "I try to understand how a technical system exactly works"
    \item  "It is enough for me to know the basic functions of a technical
system" - reversed
    \item  "I try to make full use of the capabilities of a technical system"
\end{itemize}

At the end of all their journeys, participants answered the following questions with free text input:

\begin{itemize}[itemsep=0.5pt]
    \item What was the main stressor for you during commuting?
    \item Which events influenced you during commuting ? Name up to five positive and up to five negative events.
    \item Will you continue to commute using public transport? If no: What would have to get better for you to use public transport for commuting? If yes: why?
    \item If you had to travel from Wolfenbüttel main station to Braunschweig main station (or vice versa), would you prefer using the bus or the train? Explain your answer.
\end{itemize}

\subsubsection*{Subjective measures - once per journey:}
At the beginning of every journey, participants indicated their destination and were asked to indicate every mode change of their journey by pressing the corresponding button in the MovisensXS App (Movisens GmbH in Karlsruhe, Germany, 2023b) on the study smartphone. At the end of every journey, participants rated their overall travel experience using the Satisfaction with Travel Scale (STS) \cite{ettema2011satisfaction}. The STS assesses the emotional and cognitive evaluation of the journey. It comprises of the questions 'How do you feel?' from Likert scales that measured hurried (1) to relaxed (7), worried (1) to confident (7), stressed (1) to calm (7), tired (1) to alert (7), bored (1) to enthusiastic (7) and annoyed (1) to stimulated (7). Part two of the STS asks 'How do you judge your journey?' from 'the worst journey I can imagine' (1) to 'the best I can imagine' (7), from 'low quality' (1) to 'high quality (7), and from 'the journey worked out badly' (1) to 'the trip went smoothly' (7).

\subsubsection*{Subjective measures - during travel:}
During their journey, participants were prompted to answer an experience sampling questionnaire every 3.5 minutes using the same app. The decision to adopt a 3.5-minute interval was based on a previous study \cite{luther2023, bosch_ihsed}, during which participants were exposed to different frequencies of questionnaire requests. Participants used Likert scales ranging from 1 (indicating low stress and low satisfaction) to 10 (reflecting high stress and high satisfaction) to assess their current satisfaction with the journey and their stress levels. The precise questions asked as translated from German were 'Please rate your current satisfaction with travel' and 'how stressed do you feel?'. They furthermore answered 'how do you feel independent of your journey?' from 1 (positive) to 10 (negative), 'How would you describe your current emotion?' (free text input), and 'Did anything happen since the last journey?'. If answered with 'yes', they could choose from the following event categories that were found in a prestudy \cite{luther2023, bosch_ihsed}:  

\begin{itemize}[itemsep=0.5pt]
    \item "Delay / Connection missed / Hurry required," (will be referred to as 'Delay')
    \item "Disturbing people in the environment / crowded transportation," (will be referred to as 'Disturbing People')
    \item  "Driving behavior of the driver," (will be referred to as 'Driving Behavior')
    \item  "Infrastructure (e.g. no roofing/no or not enough information /not barrier-free / uneven road)," (will be referred to as 'Infrastructure')
    \item  "Positive social situation/interaction," (will be referred to as 'Positive Social Interaction')
    \item "Media entertainment (music, podcast, book, etc.)," (will be referred to as 'Media Entertainment')
    \item  "Destination or connection reached / means of transport on time," (will be referred to as 'Destination/connection reached')
    \item  "Discomfort (boredom, impatience, difficult thoughts, hunger, fatigue, cold, wetness, etc.).," (will be referred to as 'Discomfort')
    \item  "Transportation is comfortable (cozy/pleasant temperature / not too crowded / get a seat, etc.)," (will be referred to as 'Comfort')
    \item  "Beautiful surroundings,"
    \item and an option to "report a different category." (will be referred to as 'Others')
\end{itemize}

If they chose 'report a different category', a free text input opened up. We translated the free text inputs of the 'emotion' and the 'event' free text input to English with \url{https://www.deepl.com/de/translator}. Between the prompted questionnaires, the participants were instructed to freely prompt another questionnaire whenever they encountered something that changed their evaluation or experience of the commute. 

\subsubsection*{Heart Rate Measures:} The participants' electrocardiogram was recorded once per second and extracted from an EcgMove 4 belt from Movisens (Movisens GmbH, Karlsruhe, Germany, 2022). The Movisens Data Analyzer Version 1.13.27 was used to extract heart rate and different measures of heart rate variability in time and frequency domain. The data that is read into the R-Markdown code '01\_Basics-read-translate.Rmd' stems from this Data Analyzer set to an output rate of 1Hz. Leys et al. \cite{leys2013detecting} suggest that the median absolute deviation (MAD) is a more robust measure of variability than commonly used z-scores. Therefore, the original heart rate data (column 'ecgbelt\_hr') was transformed using the median absolute deviation per participant (column 'ecgbelt\_hr\_mad') by 1) centering the values by subtracting the participant's median heart rate from all heart rate data, 2) calculating these values' median absolute deviation and 3) division of the centered values by the median absolute deviation values (see 02\_Preprocessing.Rmd of the available code). We excluded MAD heart rate values above 3 and below -3 following \cite{miller1991reaction, leys2013detecting}. The same MAD-correction was done for heart rate variability as measured by root mean squared differences and is saved in the column 'ecgbelt\_hrv\_mad'.

Separating changes in cardiac activity due to physical activity and emotion is a difficult task \cite{brouwer2018improving}. Previous research indicates that the initial phase of heart rate recovery, particularly following low-intensity activities like slow walking, occurs within the first minute after transitioning to a seated position. This rapid adjustment reflects the activation of the parasympathetic nervous system and the reduction in sympathetic activity, facilitating a return to near-resting heart rate levels \cite{cole1999heart}. Therefore, besides the complete heart rate and HRV data column, we provide a column where we excluded heart rate values during times that the ECG belt recognized active travel modes like walking or standing, that include data where participants were sitting for longer than one minute (see column 'ecgbelt\_hr\_mad\_filtered').

\subsubsection*{Context Data:} Throughout their travel, the time and location was consistently recorded using the movisensXS App and the Movinglab App (\url{https://movinglab.dlr.de/}). While the movisens XS app collected subjective and GPS data, the Movinglab App automatically recognized the travel mode (walk, stationary, train, change, bus, car, bicycle, unknown). To adhere to the guidelines set by our institute's data protection committee, participants received explicit instructions to activate the GPS data on the study cell phone solely upon entering the initial major street subsequent to commencing their journey. The GPS signal had a mean accuracy of $8 \pm 12m$.

\section*{Data Records}

%The Data Records section should be used to explain each data record associated with this work, including the repository where this information is stored, and to provide an overview of the data files and their formats. Each external data record should be cited numerically in the text of this section, for example \cite{Hao:gidmaps:2014}, and included in the main reference list as described below. A data citation should also be placed in the subsection of the Methods containing the data-collection or analytical procedure(s) used to derive the corresponding record. Providing a direct link to the dataset may also be helpful to readers (\hyperlink{https://doi.org/10.6084/m9.figshare.853801}{https://doi.org/10.6084/m9.figshare.853801}                                                               ) .

%Tables should be used to support the data records, and should clearly indicate the samples and subjects (study inputs), their provenance, and the experimental manipulations performed on each (please see 'Tables' below). They should also specify the data output resulting from each data-collection or analytical step, should these form part of the archived record.
The data is available on the Open Science Framework at \url{https://osf.io/rgkvq} under a CC-By Attribution 4.0 International license. Each participant's data file (.csv) is saved in the folder 'data.zip'. ExperienceMapsDataset.csv contains the complete dataset; ExperienceMapDataset\_over2 includes only areas where more than two participants contributed within a 100x100m region; and ExperienceMapDataset\_over2\_10sec provides a downsampled version at 0.1 Hz. As reading in the large data files often omits data from the final questionnaire only containing data at the very end, we also uploaded the original .xlsx-data files of the experience sampling that is not yet merged with the GPS and physiological data by the Movisens Data Analyzer in the folder ESM\_xlsx.zip. Find each column explained in Tables 5-11 at the end of his paper.

\section*{Technical Validation} 

%This section presents any experiments or analyses that are needed to support the technical quality of the dataset. This section may be supported by figures and tables, as needed. This is a required section; authors must present information justifying the reliability of their data.

\begin{figure}[ht]
\centering
\includegraphics[width=0.9\linewidth]{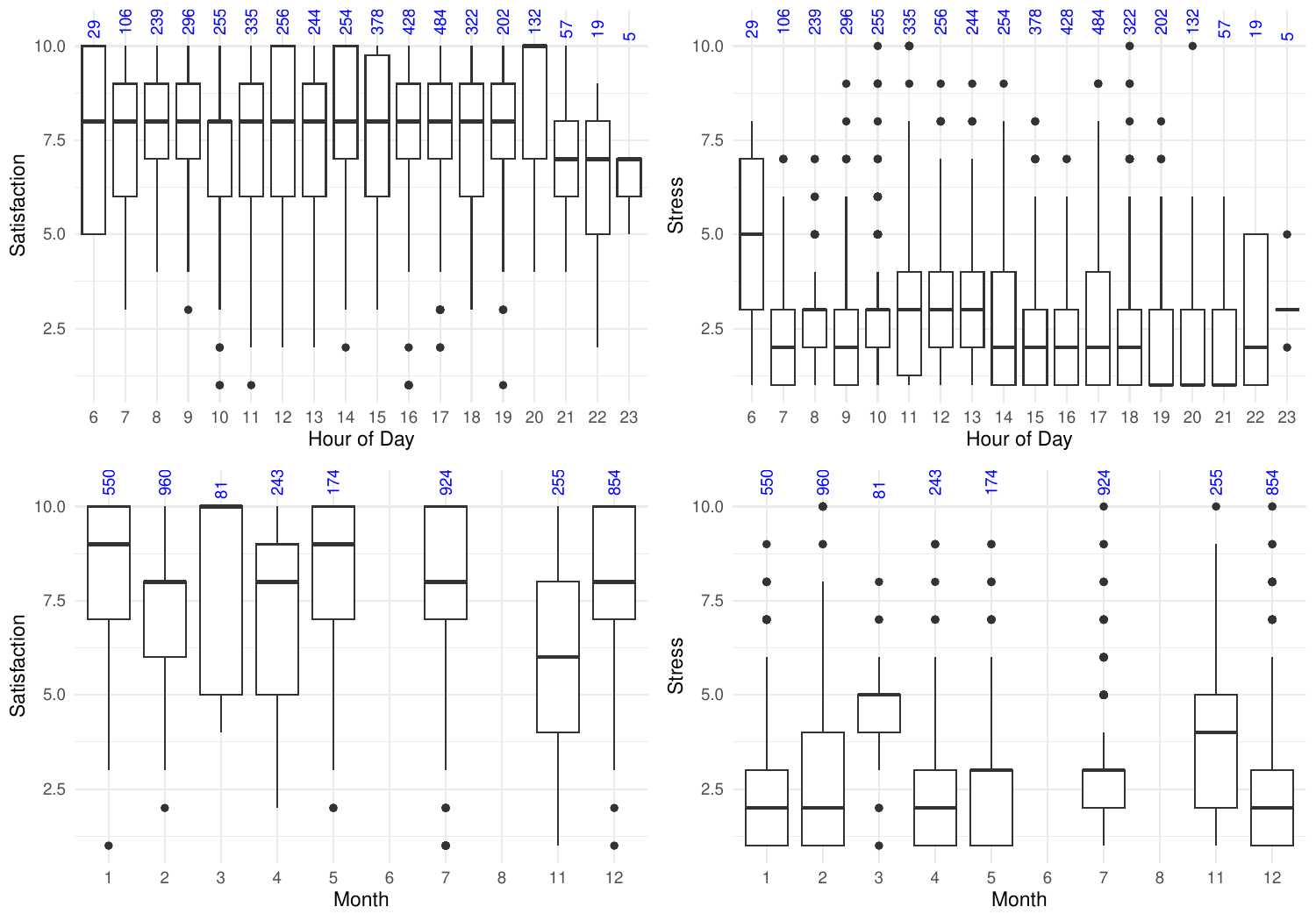}
\caption{Travel satisfaction (left) and stress (right) by day (upper plots) and year (lower plots). Blue numbers indicate the number of observations in the time bin.}
\label{seasons.}
\end{figure}

We collected 5144 ESM questionnaire answers with an average of $117 \pm 72$ answers per participant. This corresponds to an average response rate of $86.82 \pm 13.86\%$. The number of answered ESMs varied from 18 to 501 and the response rate from 50.2\% to 100\%. Participants traveled in total $547 \pm 209$ minutes over all trips as indicated by their experience sampling. The participant who traveled the shortest only traveled 261 minutes in total, the longest 1498 minutes (most likely due to a much too late press of the 'end travel' - button). Participants took at least five and at most 12 trips in total with a mean of $9 \pm 2$ trips. One trip took on average $59 \pm 26$ minutes over all participants. Figure \ref{fig:spotevents} displays the overall reported events' frequency. Figure \ref{seasons.} shows that ratings did not vary significantly across the different hours of the day or months of the year. Only  November stands out with generally lower satisfaction levels and higher stress levels. Notably, November 2023 experienced 210\% of the usual precipitation for that month in Braunschweig as indicated at \url{https://www.wetterkontor.de/de/wetter/deutschland/monatswerte-station.asp?id=10348&yr=2023&mo=-1}. In contrast, December, which had an even higher precipitation level at 287\% of the monthly average, displayed higher satisfaction levels. This discrepancy may be attributed to the nature of the precipitation: with an average temperature of 5°C in December, it is likely that much of the precipitation fell as snow rather than rain, as in November.

Figure $seasons.pdf$ goes here
Figure $EventcatsColdhotspots.pdf$ goes here

The accuracy of the GPS data was $8 \pm 12$ meters overall as indicated by the GPS measurement of the Movisens XS App.

The travel mode as reported by the participants themselves includes 101.32 hours of data collected in a bus, 48.82 hours collected in the train, 41.83 hours collected in the tram, 76.90 hours of waiting time and 50.86 hours of walking. 

\begin{figure}[ht]
\centering
\includegraphics[width=0.9\linewidth]{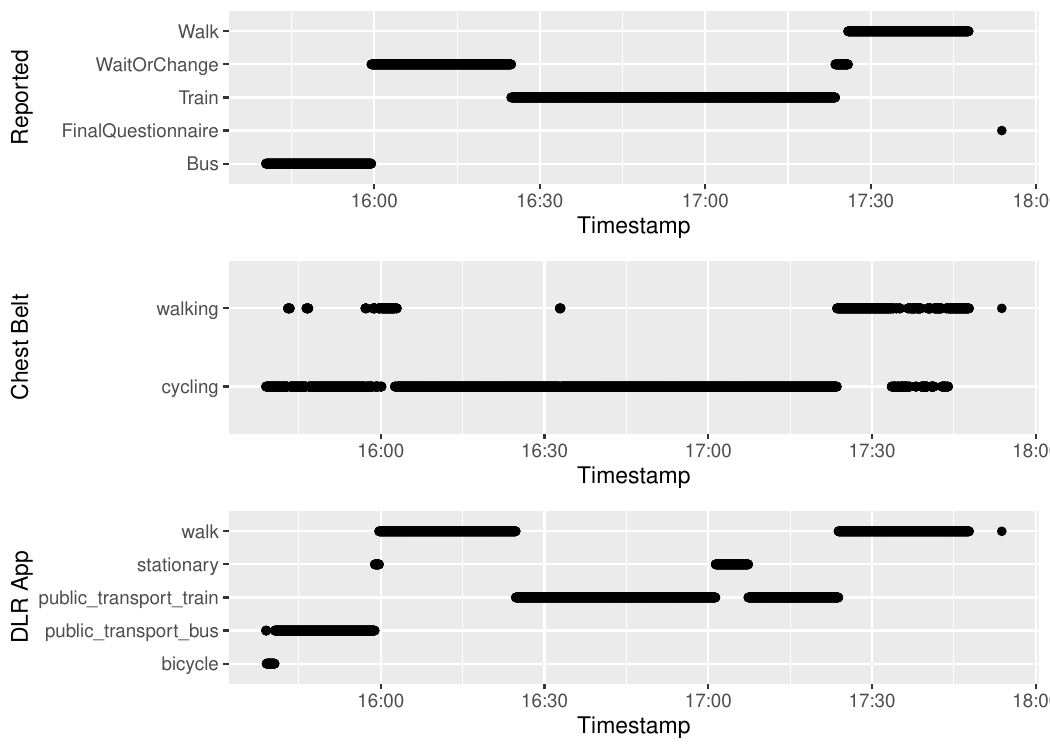}%screenshot rshiny
\caption{Travel mode of journey 1 of participant 1 as indicated by self-report, the Movisens ECG Move chest belt, and the DLR App called Keepmoving.}
\label{moderec.}
\end{figure}

\begin{figure}[ht]
\centering
\includegraphics[width=0.9\linewidth]{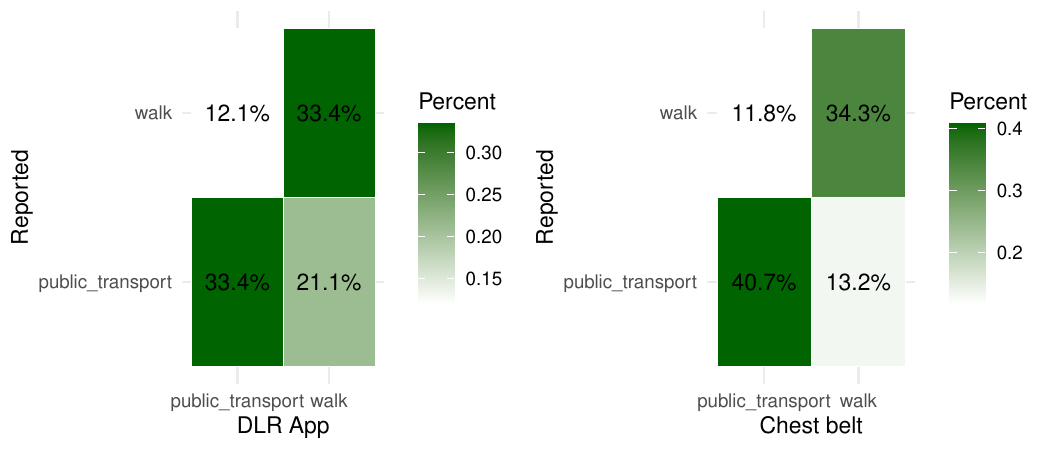}%screenshot rshiny
\caption{Confusion matrixes for travel mode detection of the DLR Keepmoving App (left) and the Movisens chest belt (right).}
\label{conf.}
\end{figure}

Self-reported travel modes agrees with the DLR Keepmoving App with a Cohen's Kappa of .711 (88\% agreement) of data points, and the Movisens ECG Move chest belt agrees with a Cohen's Kappa of .11 ( 19\% agreement) with the self-reported travel modes. Figure \ref{moderec.} displays one exemplary mode plot for journey 1 of participant 1 for self-reported mode, mode detection by the Movisens chest belt and mode detection by the DLR App movisens. Figure \ref{conf.} displays the confusion matrix between the travel mode recognition types for the mode types that they overlapped.

Figures moderec.pdf and conf.pdf go here.

\begin{figure}[ht]
\centering
\includegraphics[width=0.9\linewidth]{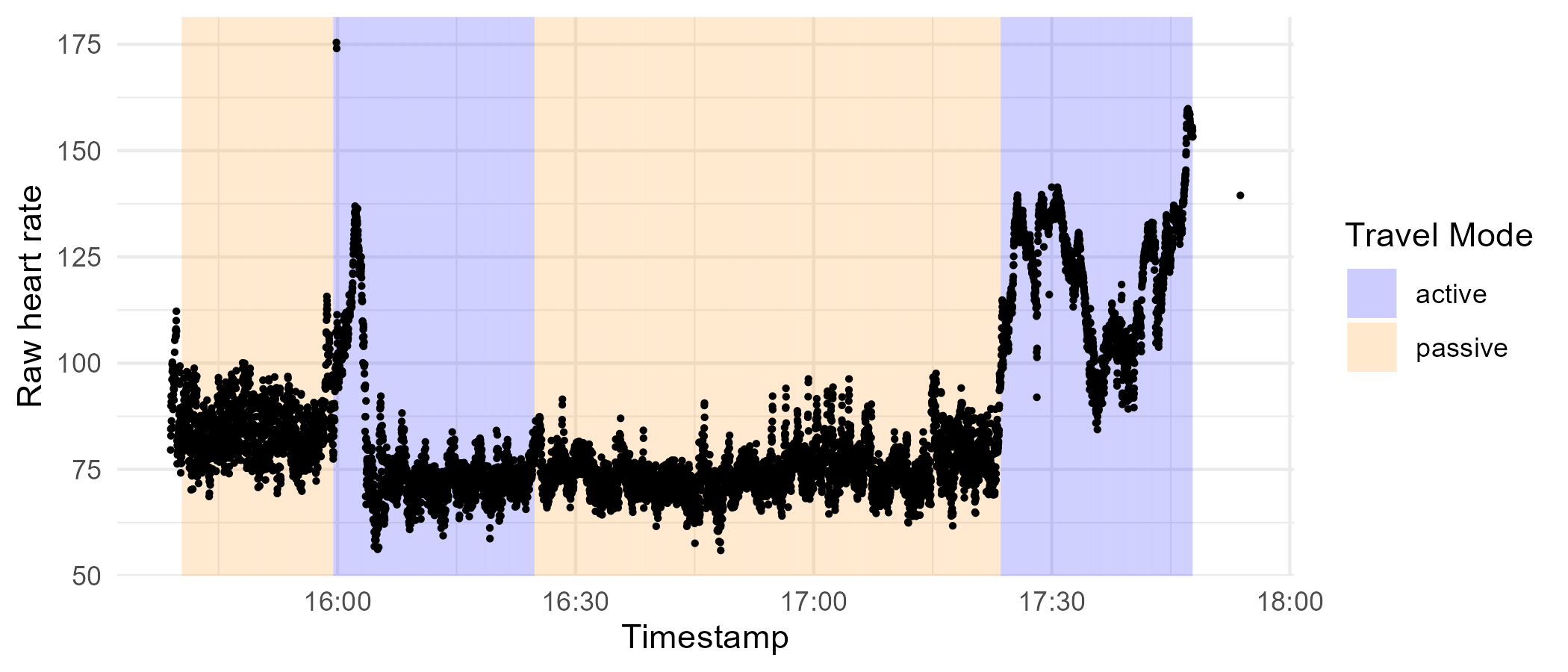}%screenshot rshiny
\caption{Raw heart rate data of journey 1 of participant 1 colored by whether the transportation mode was an active or passive travel mode.}
\label{HRactivepassive.}
\end{figure}

\begin{figure}[ht]
\centering
\includegraphics[width=0.9\linewidth]{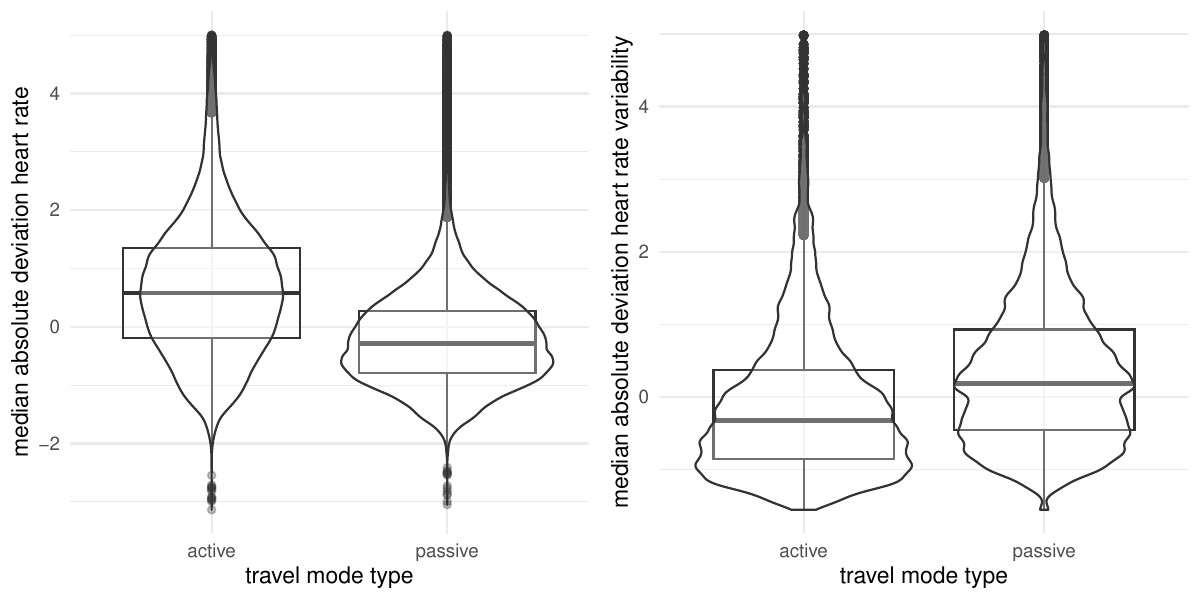}
\caption{median absolute deviation heart rate and heart rate variability in active (walk, bike, wait) and passive (train, bus tram) travel mode types.}
\label{HRactive.}
\end{figure}

The mean heart rate over all participants is $88.3 \pm 20.9$ beats per minute with a range from 30.04 to 240. The mean root mean square of successive differences (RMSSD) as measure for heart rate variability is $29.71 \pm 19.00$ over all participants.
Figure \ref{HRactivepassive.} shows an exemplary heart rate measure during journey 1 of participant 1 with a background of the variable indicating whether this was an 'active' (wait, walk, bike) or 'passive' (train, bus, or tram) transport mode.
The average MAD - heart rate during active travel modes is $0.69 \pm 1.18$ (median 0.58) compared to an average of $-0.21 \pm 1.18$ (median 0.32) in passive travel modes (see Figure \ref{HRactive.}). The average MAD - heart rate variability as measured by root mean squared error during active travel modes is $-0.10 \pm 1.03$ (median -0.33) compared to an average of $0.36 \pm 0.09$ (median 0.18) in passive travel modes (see Figure \ref{HRactive.}). The correlation between the during-travel stress rating and the MAD - heart rate is .1 with a 95\% confidence interval of .06 to .14 ($p < .001$, Cohen's $d = .20$) in the passive travel modes.

Figures $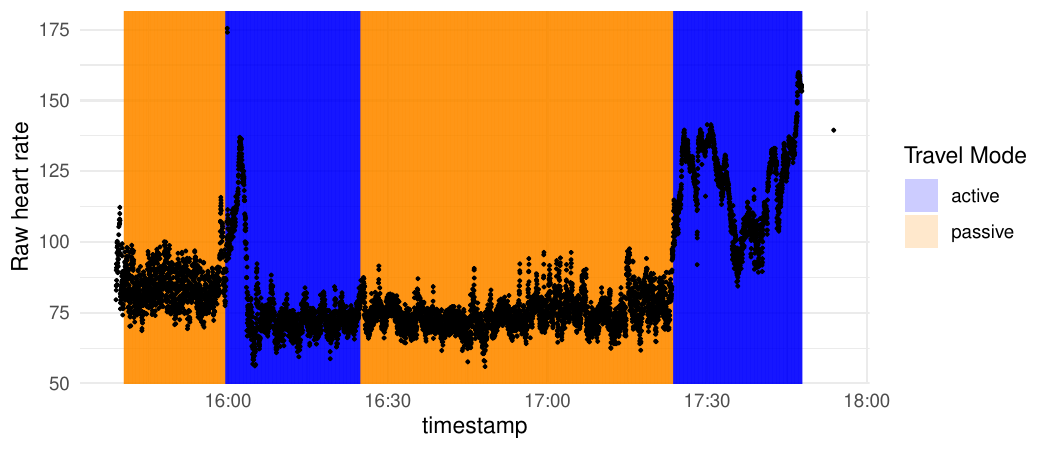$ and activeHR.pdf go here.

\begin{figure}[ht]
\centering
\includegraphics[width=0.9\linewidth]{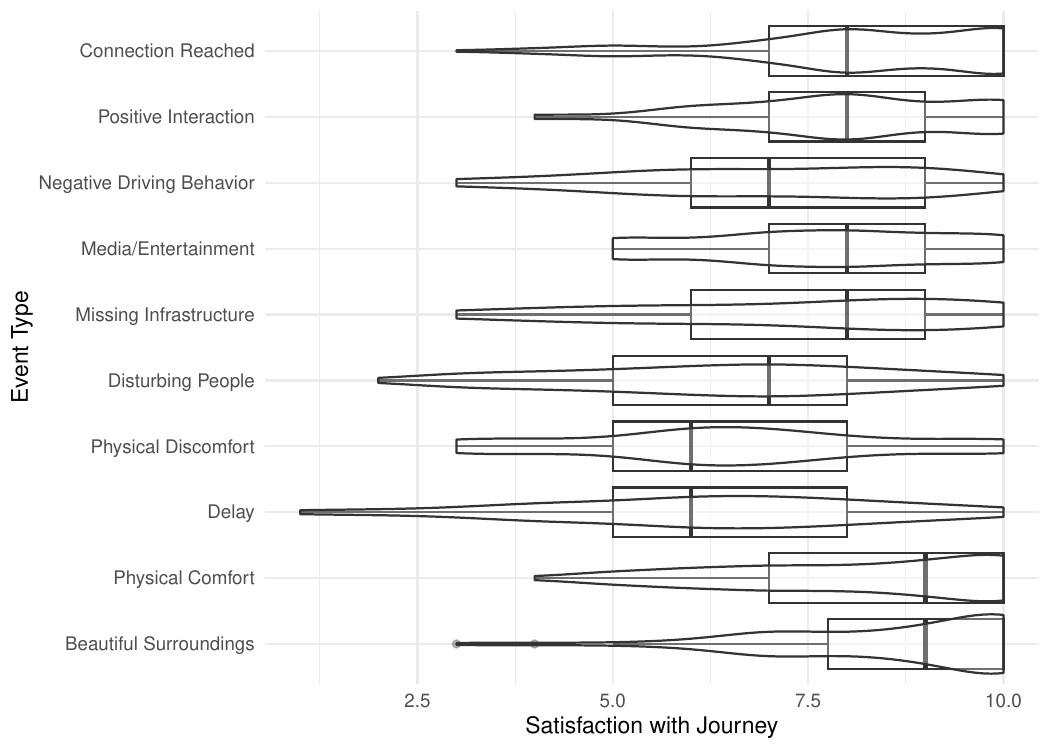}%screenshot rshiny
\caption{Travel satisfaction by event.}
\label{Travelsatisev.}
\end{figure}

The satisfaction with the journey by event is shown in Figure \ref{Travelsatisev.}. On average, the begative events 'negative driving behavior', 'missing infrastructure', 'disturbing people', 'physical discomfort' and 'delay' have an average travel satisfaction rating of $6.48 \pm 2.08$. In the positive events 'connection reached', 'positive interaction', 'media entertainment', 'physical comfort' and 'beautiful surroundings' have an average satisfaction rating of $8.14 \pm 1.68$. 

Figure eventsatis.pdf goes here.

\begin{figure}[ht]
\centering
\includegraphics[width=\linewidth]{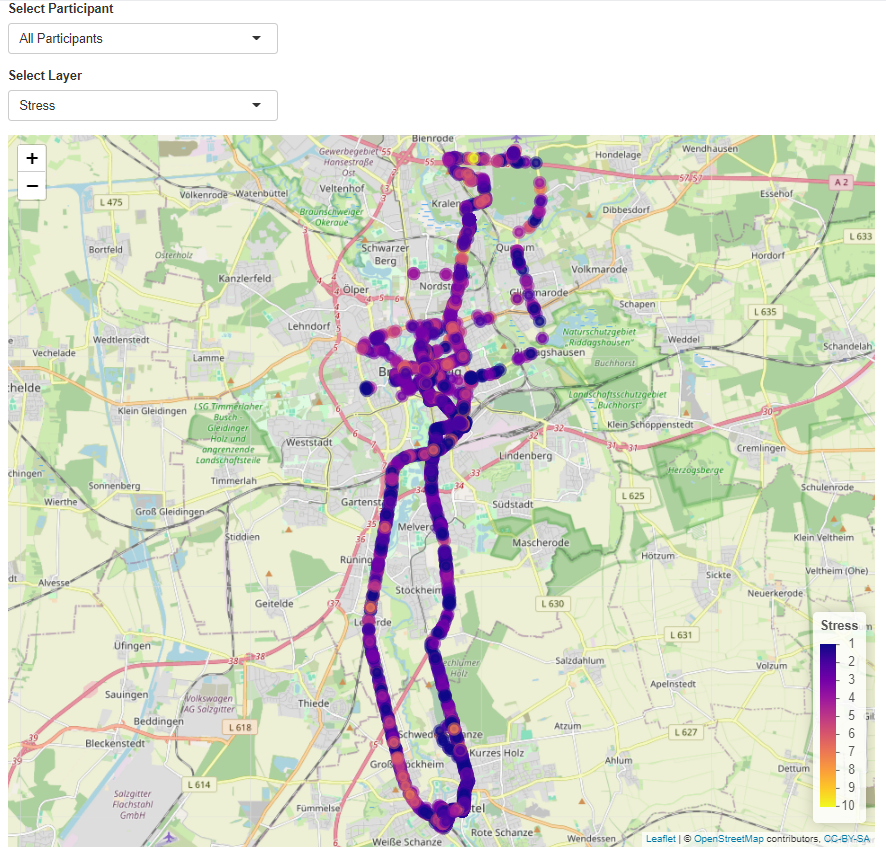}%screenshot rshiny
\caption{Screenshot of the interactive map.}
\label{shinyfigure}
\end{figure}

We include an interactive map at \url{https://estherbosch.shinyapps.io/affectivemaps} that enables dynamic visualization of our dataset with the R Shiny package \cite{shiny} in R \cite{R}. The data displayed on the map were filtered to only include data where more than two participants contributed to an area in order to deceive the participant's home street in case they started the GPS tracking earlier than instructed. This was done by rasterizing the data into 100mx100m longitude and latitude squares and excluding all squares that contained less than two participants. All relevant variables measured during the journeys can be chosen by a drop-down menu and data of all participants or single participants can be selected, see Figure \ref{shinyfigure}. A click on each data point provides all available information on this point's stress rating, satisfaction rating, heart rate, and freely reported emotions.

Figure shiny.PNG goes here.

\subsubsection*{Cold-/Hot Spots} We calculated cold spots and hot spots of satisfaction and stress ratings using the Getis-Ord Gi\* statistic following Getis \cite{getis1992analysis} with 100m radius circles around each data point. This is an analysis to indicate a statistically significant cluster of high or low values in each rating. A statistically significant hot spot exhibits a high value and has neighboring high values. The cumulative value of a feature (the data points within the circle around a data point with a 100m radius, in this case) and its adjacent data points are assessed relative to the cumulative value of all features. When this cumulative value significantly diverges from the anticipated local sum more than expected by random chance, it represents a statistically significant stress or satisfaction rating. To all calculations, we applied Bonferroni correction for multiple tests. The resulting hot spots and cold spots are visualized on an interactive map, showing all data points within the 100-meter radius and highlighting statistically significant areas. However, it is important to interpret these results cautiously given the limited number of data points. With the small sample size, the results may be overfitted or less generalizable, and the identified hot and cold spots could be influenced by outliers or limited regional variation. Data from 34 out of 44 participants contributed to the satisfaction cold spot analysis and fourteen out of 44 participants reported their stress levels within the identified stress hot spots, contributing to the calculation of these locations.

\begin{figure}[ht]
\centering
\includegraphics[width=0.6\linewidth]{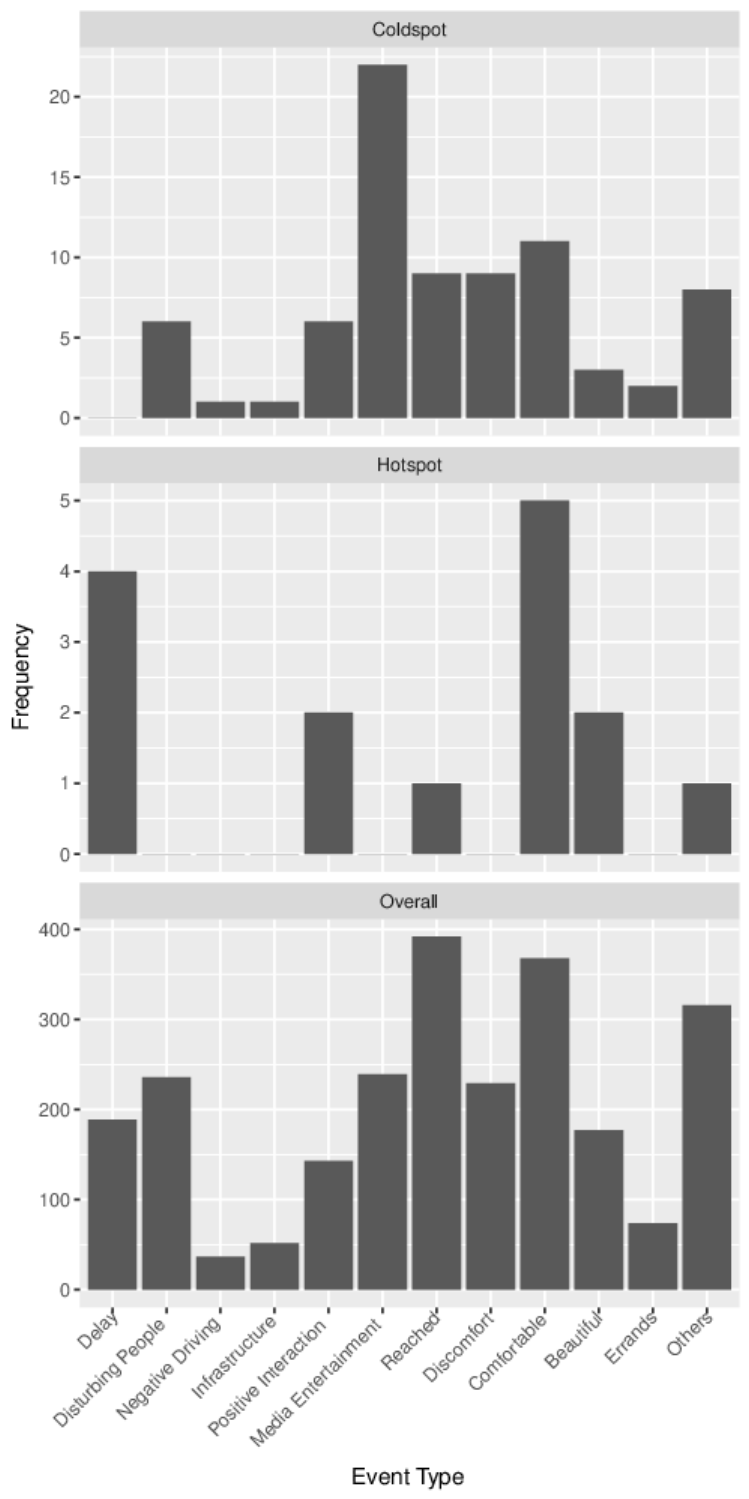}
\caption{Event categories reported  top: overall, middle: in the stress hot spot and bottom: the satisfaction cold spot.}
\label{fig:spotevents}
\end{figure}

One satisfaction cold spot was located at the main interchange station in Braunschweig, where participants were on their way from or to home. This station is often crowded, and finding the correct tram station can be challenging due to the multiple options available. The second satisfaction cold spot was found in the center of Wolfenbüttel, approximately 500 meters east of the designated start and end point of the journey. At the end of the study several participants reported frustration about having to travel all the way to Wolfenbüttel center only to immediately turn back and begin their return journey. While the 500-meter offset is unexplained, this area therefore roughly aligns with expectations as a satisfaction cold spot. No statistically significant satisfaction hot spots were identified. The reported events within the cold spot are displayed in Figure \ref{fig:spotevents}. 28 freely reported emotions in the satisfaction cold spots, of which 11 (39.3\%) were  positive as evaluated by the first author ("calm", "happy", "joy", "relaxed", "satisfied", "in a good mood and relaxed", "motivated", "excited", "good", "serene", "enjoyable"), also 11 (39.3\%) were negative ("upset", "annoyed", "tired and funny to drive past home", "tired", "slightly tense", "tense", "a little annoyed and tired", "annoyed but getting better", "still annoyed", "slightly annoyed", "bored") and 6 (21.4\%) neutral ("okay", "neutral", "indifferent", "unchanged", "slow", "quiet"). The mean satisfaction in the satisfaction cold spots is $6.71 \pm 1.60$, compared to $7.67 \pm 1.90$ in all data.

One of the stress hot spots is located along a segment of the route that falls outside the official study route, making its origin unclear. This hot spot corresponds to a street serviced only by buses, and it is likely linked to several participants' journeys, either heading to or returning from their homes. The second stress hot spot is located on the train route just before Wolfenbüttel. At this location, all participants were traveling by train. It is possible that stress levels were influenced by concerns about catching the train back while still having to walk to Wolfenbüttel city center. No stress cold spots were identified. The reported events within the hot spot are displayed in Figure \ref{fig:spotevents}. There were 12 freely reported emotions in the stress hot spots, of which 3 (25\%) were positive ("Good","Relaxed", "Satisfied"), 4 (33.33\%) were negative ("Stressed",  "Tired", "Stressed because the tracked mode does not show up","Annoyed") and 5 (41.67\%) neutral ("Quiet", "Okay", "Good, a little stressed because late, a lot to do", "Talkative", "Neutral"). The mean stress rating in the stress hot spots was $5.32 \pm 2.24$ compared to a mean stress rating of $2.70 \pm 1.70$ overall.

\subsubsection{Heart Rate Measures and Permutation Analysis of Heart Rates in Cold and Hot Spots} As reference analysis concerning the heart rate and HRV data we compared the heart rates in the overall satisfaction cold spot and stress hot spot against all other heart rates by permutation analysis. For this, we drew a sample with the same sample size as the hot and cold spot, respectively, 10,000 times. We compared the hot and cold spot MAD-transformed heart rate data to the distribution of MAD- transformed heart rates in the 10,000 samples. Please note that the limited sample size of 44 participants with 4013 valid heart rate values for the hot spot and 16591 valid heart rate values for the cold spot may reduce the reliability of the permutation-based results, potentially compromising statistical power. The heart rate data of satisfaction cold spot's heart rate data resulted in significantly lower heart rates in the satisfaction cold spot with a permutation-based p-value of $p <.001$, Cohen's $d: .09$ and an observed mean of $ -0.25$ . Similarly, the heart rate was significantly lower in the stress hot spot (permutation-based p-value: $p < .001$, Cohen's $d: .05$ , observed mean = $-0.18$ ). HRV did not show significant differences between hot- and cold spots and the rest of the data.

\section*{Usage Notes}

%The Usage Notes should contain brief instructions to assist other researchers with reuse of the data. This may include discussion of software packages that are suitable for analysing the assay data files, suggested downstream processing steps (e.g. normalization, etc.), or tips for integrating or comparing the data records with other datasets. Authors are encouraged to provide code, programs or data-processing workflows if they may help others understand or use the data. Please see our code availability policy for advice on supplying custom code alongside Data Descriptor manuscripts.

%For studies involving privacy or safety controls on public access to the data, this section should describe in detail these controls, including how authors can apply to access the data, what criteria will be used to determine who may access the data, and any limitations on data use. 

The accompanying R code reproduces all the results presented in this paper but is not necessary for further use of the dataset. The R-Markdown '01\_Basics-read-translate.Rmd' and '02\_Preprocessing.Rmd' contain all the preprocessing of the data, '03\_hotcold spots.Rmd' contains the code for finding the hot and cold spots of stress and travel satisfaction, and '04\_Rshinyapp.Rmd' contains the downsampling code and interactive visualization. 

Preprocessed cardiac activity data is already available within the provided columns. We recommend to use the interactive map to get an overview of the data. We also recommend to use Heart Rate and HRV only filtered by physical activity through transportation mode. For a more reliable determination of participants' physical activity, we suggest integrating and interpreting GPS data, self-reported transportation modes, and the modes detected by the MovingLab app. All original cardiac activity data are included in the data. The MAD-transformed values of RMSSD and heart rate are one suggestion to exclude cardiac activity data during and one minute after physically active traveling modes. Literature suggests that a sitting period of 1 to 10 minutes is sufficient for analyzing cardiac activity data after physical activity \cite{cole1999heart, shetler2001heart}. The specific duration within this range should be determined by the researcher, depending on the desired level of strictness in discarding potentially affected data. As the times of physical activity last usually at least a few minutes, we do not recommend to use any imputations. 

Note that not all Experience Sampling data have a corresponding Longitude and Latitude value, so that filtering by valid Longitude and Latitude values excludes some of the freely reported experience sampling values. This is, therefore, the case for all dataframes but the original 'ExperienceMapsDataset.csv'. For translation of the columns with free text input we recommend to use \url{https://www.deepl.com/de/translator}. As demonstrated in the technical validation of travel mode detection, none of the three travel mode columns - ModeButton (subjective reporting), Mode\_keepmoving (from the DLR app), and ActivityClass (from the chest belt) — achieved perfect accuracy. Errors occurred due to participants forgetting to indicate mode changes and instances where the app or chest belt incorrectly recognized travel modes. To address these limitations, we recommend leveraging a combination of all three detection methods. For instance, subjective mode reports can be visualized on a map and cross-validated against the plausibility of travel modes based on street types. If inconsistencies are detected, the DLR app and ECG chest belt travel mode detections can provide additional context. Furthermore, for imputing missing GPS data, the open-source routing software Valhalla offers a viable solution (\url{https://github.com/valhalla}).

To include more context data an option is to use OpenStreetMaps' export data which contains raw geographic and metadata information about nodes, ways, and relations. To obtain weather information, we recommend to use One Call API 3.3 from OpenweatherMap (\url{https://openweathermap.org/api/one-call-3}). As Figure \ref{seasons.} indicates, satisfaction levels seem to be generally lower in November than the rest of the year, and a correlation between weather and satisfaction data could be interesting.

The dataset is geographically limited, including data only from the German cities of Braunschweig and Wolfenbüttel. Consequently, the generaliezability of the findings must be further evaluated in future studies employing similar data acquisition methods. To enhance the robustness of future data collection, the approach could be expanded to include survey items inspired by datasets such as the Household, Income, and Labor Dynamics in Australia (HILDA) dataset \cite{wooden2024data, summerfield2023hilda}. While efforts were made to balance the sample by gender, participants were drawn exclusively from our institution's participant pool, which may skew the demographics toward individuals with prior experience in research studies or interested in the research topics as well as those motivated by financial incentives. Additionally, the sample's age distribution and educational or occupational diversity were not controlled, which could introduce biases in interpreting results (see Tables \ref{tab:clivingsituation}, \ref{tab:education} for demographic data). Self-selection bias of participants is more likely in experience sampling studies than other studies due to the intense time-demands on participants - and certain types of individuals might answer more frequently than others, even though incentivization by money has shown to be a good solution to this \cite{scollon2003experience}. The relatively small sample size (44 participants) represents a limitation in terms of the generalizability of the findings. The insights from this dataset should therefore be interpreted with caution, particularly when considering its applicability to broader populations. However, the sample size aligns with the exploratory nature of the study, which focuses on integrating subjective, cardiac activity, and contextual data to investigate travel experiences. Future studies should aim to replicate these methods with larger and more representative samples to validate and extend these findings. %The geographical focus on a single region in Germany and limited diversity in geographic, age, and socioeconomic backgrounds constrain the generalizability of the dataset to broader contexts. Future users of this dataset should consider these factors when analyzing or generalizing findings.

\section*{Code availability}

%For all studies using custom code in the generation or processing of datasets, a statement must be included under the heading "Code availability", indicating whether and how the code can be accessed, including any restrictions to access. This section should also include information on the versions of any software used, if relevant, and any specific variables or parameters used to generate, test, or process the current dataset. 

All code (written in R) used to process our data as explained in this paper (Preprocessing, Experience Maps, Hot and cold spots Analysis) can be found in the same folder as the dataset.

\section*{Author contributions statement}
%following CRediT Roles

Conceptualization, E.B., K.I.; 
data curation, E.B., R.L.;
formal analysis, E.B.; 
funding acquisition: K.I.; 
investigation, E.B., R.L.; 
methodology, E.B., R.L., K.I.; 
project administration, E.B.;
resources, E.B., R.L.; 
supervision, E.B.; 
validation (Event categories), R.L.; validation (Hot-cold spot Analyses), E.B.;
visualization, E.B.; 
writing—original draft, E.B.; 
All authors have read and agreed to the manuscript. This manuscript was spell-checked and rephrased using ChatGPT Version 4o in November 2024.

\section*{Competing interests} The authors declare no competing interests.

\newpage
\section*{Figures \& Tables} %screenshot, shinyapplink

%Figures, tables, and their legends, should be included at the end of the document. Figures and tables can be referenced in \LaTeX{} using the ref command, e.g. Figure \ref{fig:stream} and Table \ref{tab:example}. 

%Authors are encouraged to provide one or more tables that provide basic information on the main ‘inputs’ to the study (e.g. samples, participants, or information sources) and the main data outputs of the study. Tables in the manuscript should generally not be used to present primary data (i.e. measurements). Tables containing primary data should be submitted to an appropriate data repository.

%Tables may be provided within the \LaTeX{} document or as separate files (tab-delimited text or Excel files). Legends, where needed, should be included here. Generally, a Data Descriptor should have fewer than ten Tables, but more may be allowed when needed. Tables may be of any size, but only Tables which fit onto a single printed page will be included in the PDF version of the article (up to a maximum of three). 

%Due to typesetting constraints, tables that do not fit onto a single A4 page cannot be included in the PDF version of the article and will be made available in the online version only. Any such tables must be labelled in the text as ‘Online-only’ tables and numbered separately from the main table list e.g. ‘Table 1, Table 2, Online-only Table 1’ etc.

% Commute Distances
\begin{table}[ht]
\centering
\begin{tabular}{lcccccccc}
\toprule
\textbf{Distance} & \textbf{<1 km} & \textbf{1-5 km} & \textbf{5-10 km} & \textbf{10-20 km} & \textbf{20-30 km} & \textbf{30-40 km} & \textbf{40-50 km} & \textbf{>50 km} \\
\midrule
Count & 4 & 20 & 10 & 5 & 3 & 1 & 1 & 0 \\
\bottomrule
\end{tabular}
\caption{Participants' Commute Distances.}
\label{tab:commutedistance}
\end{table}

% Living Situations
\begin{table}[ht]
\centering
\begin{tabular}{lc}
\toprule
\textbf{Living Situation} & \textbf{Count} \\
\midrule
Countryside            & 1  \\
Up to 10,000           & 0  \\
Up to 50,000           & 4  \\
Up to 100,000          & 3  \\
Up to 500,000          & 35 \\
Larger than 500,000    & 0  \\
\bottomrule
\end{tabular}
\caption{Participants' Living Situation.}
\label{tab:clivingsituation}
\end{table}

\vspace{1em}

% Education Levels
\begin{table}[ht]
\centering
\begin{tabular}{lc}
\toprule
\textbf{Education Level} & \textbf{Count} \\
\midrule
None                  & 1  \\
Primary School        & 0  \\
Secondary School      & 4  \\
Technical High School & 2  \\
High School           & 21 \\
University            & 16 \\
\bottomrule
\end{tabular}
\caption{Participants' Education Levels.}
\label{tab:education}
\end{table}

\vspace{1em}

\begin{table}[ht]
\centering
\begin{tabular}{lcccccccc}
\toprule
\textbf{Commute Mode} & \textbf{Walk} & \textbf{Bike} & \textbf{Bus} & \textbf{Train} & \textbf{Driver} & \textbf{Co-driver} & \textbf{Tram} & \textbf{Other} \\
\midrule
Count & 18 & 24 & 29 & 11 & 9 & 1 & 16 & 2 \\
\bottomrule
\end{tabular}
\caption{Participants' Commute Modes.}
\label{tab:commutemode}
\end{table}

%\begin{figure}[ht]
%\centering
%\includegraphics[scale=0.8]{BFI.pdf}
%\caption{Answers to the Big Five Inventory (BFI-10) following %\cite{rammstedt2013short}.}
%\label{fig:bfi}
%\end{figure}

%\begin{figure}[ht]
%\centering
%\includegraphics[scale=0.8]{ATI.pdf}
%\caption{Answers to the Affinity to Technology (ATI-9) scale %following \cite{franke2019personal}.}
%\label{fig:ati}
%\end{figure}

%\begin{figure}[ht]
%\centering
%\includegraphics[width=0.9\linewidth]{travelmodetime.pdf}%screenshot rshiny
%\caption{Total duration of collected data for each travel mode.}
%\label{modetime.}
%\end{figure}

\begin{table}[ht]
\begin{tabular}{p{0.2\textwidth}|p{0.5\textwidth}|p{0.2\textwidth}}
\hline
\textbf{Column} &
  \textbf{Content} &
  \textbf{Value Range} \\ \hline
timestamp &
  the data point's time stamp in the format &
  the whole year 2023 \\ \hline
participant &
  participant number &
  1-44 \\ \hline
age &
  participant's age &
  free text input
   \\ \hline
gender &
  participant's self-identified gender &
  1: male; 2: female; 3: diverse
   \\ \hline
education &
  participant's highest educational degree &
  1: none; 2: primary school; 3: secondary school; 4: technical high school; 5: high school; 6: university
   \\ \hline
city &
  What is your current living situation? &
  1: countryside; 2: city up to 10000 inhabitants; 3: up to 50000; 4: up to 100000; 5: 500000; 6: larger than 500000 \\ \hline
commute &
  How far is your commute to work in kilometers? &
  1: less than 1 km; 2: 1-5km; 3: 5-10km; 4: 10-20; in steps of 10km until 8: over 50km. \\ \hline
cardiac &
  whether the participant had any cardiac diseases &
  1 if true; 2 if not
   \\ \hline
cardiac\_free &
  description of the cardiac disease &
  free text input
   \\ \hline
commute\_ &
  all columns that start with Commute indicate the percentage of the travel mode used usually for commuting to work &
  0-100
   \\ \hline
  BFI\_reserved &
  all columns that start with BFI indicate the questions of the big five inventory as explained in the measurements section &
  1: not at all; 5: fully applies.
   \\ \hline
ATI\_detail &
  all columns that start with ATI indicate the questions of the affinity for technology inventory as explained in the measurements section &
  1: not true at all; 6: completely correct.
   \\ \hline
license &
  whether the participant has a driver's license &
  1: yes, 2: no
   \\ \hline
week\_car &
  all columns that start with week indicate the percentage of the travel mode usually used during the week & 0 to 100
   \\ \hline
weekend\_car &
  all columns that start with weekend indicate the percentage of the travel mode usually used during weekends & 0 to 100
   \\ \hline
license\_since &
  year since when the participant has a license &
  free number input
   \\ \hline
Drove\_car\_1 &
  all columns that start with Drove indicate how many km per year the participant usually drives. &
  1: below 5000km; 2: 5000-10000; 3 to 7: steps of 10000km; 8: over 60000km. \\ \hline
  bus\_vs\_train &
  If you had to travel from Wolfenbüttel main station to Braunschweig main station (or vice versa), would you prefer using the bus or the train? &
  1: bus; 2: train \\ \hline
arg\_bus\_vs\_train &
  What are the reasons for your answer? &
  free text input \\ \hline  
biggest\_stressor &
  What was the main stressor for you during commuting? &
free text input \\ \hline
wish\_stressor &
  What would you have wished for in that stressful situation? &
  free text input \\ \hline
  
continue\_commute &
  Will you continue to commute using public transport? &
  1: yes; 2: no \\ \hline
pro\_pt &
  if they chose yes to the previous question: why will you continue using public transport? &
  free text input \\ \hline
needed\_improvs &
  if they chose no to the previous question: What would have to improve so that you choose public transport? &
  free text input \\ \hline
pos\_1 to pos\_5 &
  Which events influenced you during commuting? Name up to five positive and up to five negative events. &
  1: yes; 2: no \\ \hline
intensity\_pos\_1 to intensity\_pos\_5 &
  How did this event influence your comfort? &
  1: very little; 5. very strongly \\ \hline

\end{tabular}%
\caption{Columns names, their content and value ranges for the variables that were collected once per participant.}
\label{tab:columns1}
\end{table}

\begin{table}[ht]
\begin{tabular}{p{0.2\textwidth}|p{0.5\textwidth}|p{0.2\textwidth}}
\hline
\textbf{Column} &
  \textbf{Content} &
  \textbf{Value Range} \\ \hline

Train\_Destination &
  Destination of the train ride &
  free text input \\ \hline
Bus\_nr &
  Bus number of this journey &
  free text input \\ \hline
Bus\_Destination &
  Destination of the bus ride &
  free text input \\ \hline
Tram\_nr &
  Tram number of this journey &
  free text input \\ \hline
Tram\_destination &
  destination of the tram journey &
  free text input \\ \hline
Restr\_Luggage &
  all columns that start with Restr indicate restrictions the participant had on this journey &
  1 if true
   \\ \hline
   journey &
  this button was pressed when the journey was done &
  1 when button was pressed
   \\ \hline
Route\_knowledge &
  How well the participant knew the journey's route &
  1: not at all; 5: very well
   \\ \hline
sts\_1 &
  all columns starting with sts indicate the questions of the satisfaction with travel scale answered after every journey was done, as described in ref methods &
  from 1 to 7 according to the satisfaction with travel scale items, respectively
   \\ \hline
\end{tabular}%
\caption{Columns names, their content and value ranges for all variables that were collected once per journey.}
\label{tab:columns2}
\end{table}

\begin{table}[ht]
\begin{tabular}{p{0.2\textwidth}|p{0.5\textwidth}|p{0.2\textwidth}}
\hline
\textbf{Column} &
  \textbf{Content} &
  \textbf{Value Range} \\ \hline

%physical activity
ecgbelt\_moveacc & from \url{https://docs.movisens.com/}: 'Movement Acceleration (aka Movement Acceleration Intensity, MAI) is a physical activity metric that outputs values that have a very good correlation to the intensity of bodily movements. For daily activities good sensor positions are at the participant's center of gravity (hip, chest). For each output interval (default is 1min) the values aggregated by calculating the mean value of all samples in the interval. The unit of the output parameter Movement Acceleration Intensity is [g], i.e. multiples of earth gravity (1g = 9,81 m/s2). The algorithm uses raw 3D acceleration signals from the acceleration sensor. The signals from each axis is bandpass filtered (Butterworth filter 0.25-11Hz, order 4) to remove the offset due to the acceleration of gravity and to remove high frequency content that does not directly relate to movements. From the resulting signals for each sample the vector magnitude is calculated:
\[
ma_{i} = \sqrt{ax^{2}_{bp} + ay^{2}_{bp} + az^{2}_{bp}}
\]
For each output interval (default is 1min) the vector magnitude is aggregated by calculating the mean value of all samples in the interval. The unit of the output parameter Movement Acceleration Intensity is [g], i.e. multiples of earth gravity (1g = 9,81 m/s2).' &
  [0,1.60] \\ \hline
  
ecgbelt\_nonwear &
  from the Movisens Documentation: 'This uses the acceleration data to reveal the times when the participant has removed the sensor. Internally wear time is calculated in 30s intervals. During charging the output always is Not Worn.' &
  0: not worn; 1: worn \\ \hline

ecgbelt\_incldown, ecgbelt\_inclforward, ecgbelt\_inclright &
  from the Movisens Documentation: 'This parameter displays the inclination of the body axes at the sensor location against the vertical. It calculates the mean inclinations of the three body axes from the acceleration signal and displays the value for each inclination in degrees. The values range from 0° to 180°. Internally the inclination calculated in intervals of 4s.' &
  [0.28,179.92], [0.078,178.95], [2.68,177.30] \\ \hline

ecgbelt\_bodypos &
  from the Movisens Documentation: 'The DataAnalyzer uses the inclination obtained from the acceleration signals to classify the body position.' &
  Unknown: 0; Lying supine: 1; Lying left: 2; Lying prone: 3; Lying right: 4; Upright: 5 \\ \hline

ecgbelt\_activity &
  from the Movisens Documentation: 'This algorithm detects the activity class that’s occurring during each output interval using a combination of features calculated from the accelerometer data and the barometric air pressure. A White-Box decision tree was used to determine the activity class.' [...] 'Literature and Validation: \cite{anastasopoulou2012classification}' &
  Unknown: 0; Lying: 1; Sitting/standing: 2; Cycling: 3; Jogging: 5; Walking: 7  \\ \hline
\end{tabular}%
\caption{Columns names, their content and value ranges for all variables that were collected throughout the whole journey by the chest-worn EcgMove4 sensor regarding physical activity, part 1.}
\label{tab:columns3}
\end{table}

\begin{table}[ht]
\begin{tabular}{p{0.2\textwidth}|p{0.5\textwidth}|p{0.2\textwidth}}
\hline
\textbf{Column} &
  \textbf{Content} &
  \textbf{Value Range} \\ \hline
  %stepcount
ecgbelt\_stepcount &
  from the Movisens Documentation: 'This parameter uses the acceleration signal to calculate the number of steps taken in a given output interval. At first the magnitude of the acceleration vector is calculated. After applying a low pass filter steps are detected by inspecting low and hing points while considering plausibility criteria. A minimum sequence of at least three consecutive steps is needed to allow step detection. The output interval determines how frequently StepsCount is output by summing all steps if each interval. Literature and Validation: \cite{anastasopoulou2013comparison}' &
  [0,148] \\ \hline
  
ecgbelt\_altitude &
from the Movisens Documentation: 'Using the signal from the barometric air pressure sensor, this parameter calculates the altitude above sea level. The altitude uses the barometric formula (exponential atmosphere) to reach its results, and thus, changes of air pressure due to changes in the weather can influence the output. The unit is meters [m]. The basis for measuring the change in altitude by means of a barometric pressure sensor is the fact that the atmosphere becomes denser at lower altitudes and thus the air pressure decreases with increasing altitude.
\[
H = \frac{T_{M,0}}{L_{M,0}} \cdot \left( \frac{p^{R \cdot L_{M,0}}}{P_0} - 1 \right)
\]
\[
P_0 = \text{Air pressure at Sea Level}
\]
\[
T_{M,0} = \text{Temperature at Sea Level}
\]
\[
L_{M,0} = \text{Temperature gradient}
\]
With this method, a relative height change can be measured with a resolution of up to 10 cm; statements about the absolute height (e.g. required for localization), are not possible due to the variations in air pressure due to weather conditions. The output interval determines how frequently altitude is output by taking the mean value' &
  [-139.01, 526.65]  \\ \hline
ecgbelt\_stressindex &
  from the Movisens Documentation: 'This parameter was developed by R.M. Baevsky. It is based on frequency distribution of the heart beat intervals. It has no unit.'[...]'Literatur:\cite{baevsky1997noninvasive, baevsky1999methodik, bayevsky2002hrv}' &
  [1, 9959.88] \\ \hline

ecgbelt\_speed &
  from the Movisens Documentation: 'This parameter calculates the mean of the vertical speed for each output interval. A positive value represents upward movement. The unit is [m/s] and the calculation uses the altitude calculated from barometric air pressure data.' &
  [-0.90,0.64] \\ \hline

\end{tabular}%
\caption{Columns names, their content and value ranges for all variables that were collected throughout the whole journey by the chest-worn EcgMove4 sensor regarding physical activity, part 2.}
\label{tab:columns4}
\end{table}

\begin{table}[ht]
\begin{tabular}{p{0.2\textwidth}|p{0.5\textwidth}|p{0.2\textwidth}}
\hline
\textbf{Column} &
  \textbf{Content} &
  \textbf{Value Range} \\ \hline
  
ecgbelt\_hr & from the Movisens Documentation: 'This parameter caluclates the mean heart rate per output interval. The unit is [bpm]'& [30.04,240]\\ \hline

ecgbelt\_hrvhf & High Frequency HF (HrvHf): 'This parameter calculates the mean value of the HRV parameter High Frequency (HF) per output interval. HF describes the spectral power in the frequency band between 0.15 and 0.4 Hz. The unit is [ms2].'from the Movisens Documentation& [0.14, 19614.95]\\ \hline

ecgbelt\_hrvlf & Low Frequency IF (HrvLf): 'This parameter calculates the mean value of the HRV parameter Low Frequency (LF) for each output interval. LF describes the spectral power in the frequency band between 0.04 and 0.15 Hz. The unit is [ms2].'from the Movisens Documentation&[0.5,151671.1]\\ \hline

ecgbelt\_hrvlfhf & HRV parameter Low to High Frequency Ratio LF/HF (HrvLfHf): 'HRV parameter Low to High Frequency Ratio (LF/HF) This parameter calculates the mean value of the HRV parameter Low Frequency to High Frequency Ratio (LF/HF) per output interval.' from the Movisens Documentation&[0,48.39]\\ \hline

ecgbelt\_hrvpnn50 &  HRV parameter pNN50 (HrvPnn50): 'This parameter calculates the mean value of the HRV parameter pNN50 per output interval. pNN50 is the percentage of NN intervals greater than 50ms. The unit is [\%].'from the Movisens Documentation&[0,84.13]\\ \hline

ecgbelt\_rmssd &  HRV parameter RMSSD (HrvRmssd): 'This parameter calculates the mean value of the HRV parameter RMSSD per output interval. RMSSD is the root mean square of successive differences of beat intervals. The unit is [ms].' from the Movisens Documentation&[1,193.99]\\ \hline

ecgbelt\_hrvsd1, ecgbelt\_hrvsd2, ecgbelt\_hrvsd2sd1 & HRV parameter SD1, SD2, SD1/SD2 (HrvSd1, HrvSd2, HrvSd2Sd1): 'SD1 and SD2 represent short and long term variability repectively. SD1 and SD2 can be derived from the Poincaré plot of the beat intervals. SD1 is the short half axis, SD2 is the long half axis of a fitted ellipse. SD1 and SD have the unit [ms]. Also the ratio SD2/SD1 can be calculated.' from the Movisens Documentation&[1,137.55] [1,158.83] [0.55,9.71]\\ \hline

ecgbelt\_hrvsdnn & 'This parameter calculates the mean value of the HRV parameter SDNN per output interval. SDNN describes the standard deviation of the beat intervals. The unit is [ms].' from the Movisens Documentation&[1,134.47]\\ \hline

  \end{tabular}%
\caption{Columns names, their content and value ranges for all variables that were collected throughout the whole journey by the chest-worn EcgMove4 sensor regarding ECG data. For a detailed explanation of the ECG data preprocessing, please refer to the page Algorithms - ECG, heart rate and heart rate variability from the Movisens Documentation.}
\label{tab:columns5}
\end{table}

\begin{table}[ht]
\begin{tabular}{p{0.2\textwidth}|p{0.5\textwidth}|p{0.2\textwidth}}
\hline
\textbf{Column} &
  \textbf{Content} &
  \textbf{Value Range} \\ \hline

ModeButton &
  Participant indicated a travel mode change &
  1 when the button was pressed
   \\ \hline
Start\_End & start and end of journey as indicated by participant
   & Start\_Travel, End\_Travel 
   \\ \hline
satisfaction\_journey &
  current satisfaction with the journey: Please rate your current satisfaction with travel &
  from 1 (indicating low stress and low satisfaction) to 10 (reflecting high stress and high satisfaction) \\ \hline
feeling\_independent &
  current feeling independent of journey: how do you feel independent of your journey?' &
  from 1 (positive) to 10 (negative) \\ \hline
stress &
  current stress level: how stressed do you feel? &
  from 1 (indicating low stress and low satisfaction) to 10 (reflecting high stress and high satisfaction) \\ \hline
emotion\_open &
  How would you describe your current emotion? &
  free text input
   \\ \hline
Event\_Delay &
  all Columns that start with Event\_ are the reported event categories experienced during traveling. They refer to the event categories explained in the measurements section &
  1 when the button was pressed
   \\ \hline
Event\_Free &
  Participants were free to report any events other than the given categories &
  free text input
   \\ \hline

satisfaction\_car &
  when participants indicated 'car' as travel mode, they did not received prompts every 3.5 minutes, but were asked to fill in their satisfaction with the journey after they indicated to not drive anymore &
  1: low satisfaction; 10: high satisfaction
   \\ \hline
feeling\_car &
  how the participant was feeling independent of the journey after the car drive was done &
  1: negative; 10: positive
   \\ \hline
stress\_car &
  how stressed the participant felt after the car drive was done &
  1: not at all; 10: very much
   \\ \hline
event\_car &
  did any event occur during the car drive? &
  1: yes; 2: no
   \\ \hline
event\_descr\_car &
  Description of any events that occurred during the car drive & free text input
   \\ \hline
bike\_satisfaction &
  when participants indicated 'bike' as travel mode, they did not received prompts every 3.5 minutes, but were asked to fill in their satisfaction with the journey after they indicated to not drive anymore. The other columns starting with 'bike' correspond to the 'car' post-drive questions. &
  satisfaction: 1: low satisfaction; 10: high satisfaction; 
  feeling: 1: negative; 10: positive; 
  stress: 1: not at all; 10: very much
   \\ \hline
Accuracy &
  Accuracy of the GPS data as measured by the Movisens XS App in meters &
  3.20m to 304.69m
   \\ \hline

\end{tabular}%
\caption{Columns names, their content and value ranges for all variables that were collected throughout the whole journey regarding experience sampling. All Columns that end with  originate from the Movisens experience sampling App.}
\label{tab:columns6}
\end{table}

\begin{table}[ht]
\begin{tabular}{p{0.2\textwidth}|p{0.5\textwidth}|p{0.2\textwidth}}
\hline
\textbf{Column} &
  \textbf{Content} &
  \textbf{Value Range} \\ \hline
  
ecgbelt\_temp &
  from the Movisens Documentation: 'This calculates the mean value of the sensor temperature per output interval. When the sensor is worn the sensor temperature is substantially determined by the body temperature and the ambient temperature. The unit is [°C].' &
  [13.8,36.4] \\ \hline

ecgbelt\_temp2 &
  from the Movisens Documentation: 'When the sensor is worn the sensor temperature is substantially determined by the body temperature and the ambient temperature. The unit is [°C].' &
  [13.9,36.4] \\ \hline
Mode\_keepmoving &
  travel mode as recognized by the Movinglab app &
  walk, stationary, train, change, bus, car, bicycle, unknown
   \\ \hline
ecgbelt\_hr\_mad &
  median-absolute deviation heart rate values &
  -3 to 3
   \\ \hline
ecgbelt\_rmssd\_mad &
  median-absolute deviation Root Mean Square of Successive Differences values &
  -3 to 3
   \\ \hline
ecgbelt\_hr\_mad\_filtered &
    median-absolute deviation heart rate values filtered by only included during passive activity class (sitting) for longer than 1 minute &
    -3 to 3
    \\ \hline
ecgbelt\_rmssd\_mad\_filtered &
    median-absolute Root Mean Square of Successive Differences values filtered by only included during passive activity class (sitting) for longer than 1 minute &
    -3 to 3
    \\ \hline
Longitude, Latitude &
  Longitude and Latitude as measured by the Movisens XS App &
  Longitude: 10.49217 to 10.57083; Latitude: 52.15681 to 52.31621
   \\ \hline

  \end{tabular}%
\caption{Columns names, their content and value ranges for variables that were collected by the chest-worn EcgMove4 sensor regarding temperature, variables that contain pre-processed ECG-related data, and GPS-related variables.}
\label{tab:columns7}
\end{table}

\end{document}